\documentclass[showpacs,preprintnumbers]{revtex4}
\usepackage{amsmath}
\usepackage{graphicx}

\input{tcilatex}

\begin{document}

\title{Resonant effects in the strongly driven phase-biased Cooper-pair box}
\author{S.N. Shevchenko and A.N. Omelyanchouk}
\affiliation{B. Verkin Institute for Low Temperature Physics and Engineering, 47 Lenin
Ave., 61103, Kharkov, Ukraine.}
\date{\today}

\begin{abstract}
We study the time-averaged upper level occupation probability in a strongly
driven two level system, particularly its dependence on the driving
amplitude $x_{0}$ and frequency $\omega $ and the energy level separation $%
\Delta E$. In contrast to the case of weak driving ($x_{0}\ll \Delta E$),
when the positions of the resonances almost do not depend on $x_{0}$, in the
case of the strong diving ($x_{0}\sim \Delta E$) their positions are
strongly amplitude-dependent. We study these resonances in the concrete
system -- the strongly driven phase-biased Cooper-pair box, which is
considered to be weakly coupled to the tank circuit.
\end{abstract}

\maketitle

Several mesoscopic superconducting devices, which behave as
quantum-mechanical two level systems (TLSs), were proposed and studied
recently (see reviews \cite{MShSh, WenShum}). And although these devices are
formally analogous to microscopic TLSs (such as electrons, atoms, photons,
etc. \cite{CohTan}), they differ in that the coupling to controlling gates
and an environment must be taken into account (this makes the numerical
analysis of a mesoscopic TLS necessary). The study of the dynamic behaviour
of the mesoscopic superconducting structures is interesting because they are
suitable to observe the quantum-mechanical features by measuring macroscopic
values and because of their relevance for engineering on the mesoscopic
scale, e.g. for potentially realizable quantum computers based on
superconducting Josephson qubits. The following non-stationary effects were
studied in the superconducting effectively TLSs: Rabi oscillations \cite%
{Nakamura99, Martinis02, Ilichev03, Claudon04}, multiphoton excitations \cite%
{nak01, Wallraff03, yaponci, multiphoton}, Landau-Zener transition \cite%
{Izmalkov04, Ithier}, nonlinear excitations \cite{Goorden}. In this work we
study the strongly driven superconducting TLS. Namely, we study the
phase-biased Cooper-pair box (PBCPB) (also called the Cooper-pair
transistor) \cite{Tinkham, italyanci, Zorin(02), FriedmanAverin, Krech02}
strongly driven via the gate electrode and probed by the classical resonant
contour (tank circuit). The particular interest in this problem is because
due to the interference between the Landau-Zener tunneling events, the
system can be resonantly excited and the probability of the excitation
quasi-oscillatory depends on the amplitude of the driving parameter \cite%
{TerNak, ShIF, SaiKa, ShKOK}. That is why we are interested in the dynamics
of the strongly driven superconducting TLS -- to clarify this problem and to
relate it to the experimental results \cite{VI}.

The rest of the paper is organized as following. First we analyse the
resonant excitations of a TLS, particularly, the difference between the
weakly and strongly driven regimes. Then we study concrete situation of the
strongly driven PBCPB, which is probed by the tank circuit. The paper ends
with the conclusions.

We consider a TLS described by the Hamiltonian 
\begin{equation}
\widehat{H}(t)=\Delta \widehat{\sigma }_{x}+(x_{off}+x_{0}\sin \omega t)%
\widehat{\sigma }_{z}.\text{ }  \label{Ham_ShIF}
\end{equation}%
Here $\widehat{\sigma }_{x,z}$ are the Pauli matrices. We are interested in
the time-averaged upper level occupation probability, which is assumed to be
related with the observable values. A driven TLS can be resonantly excited
from the ground state to the upper state \cite{GrifHang}. When the driving
amplitude $x_{0}$ is small compared to the energy level separation $\Delta
E=2\sqrt{\Delta ^{2}+x_{off}^{2}}$, the positions of the resonances in the
time-averaged upper level occupation probability is determined by the
multiphoton relation, $\Delta E=K\cdot \hbar \omega $. Here $\omega $ is the
driving frequency and $K$ is an integer. If the amplitude $x_{0}$ is
increased, the position of the resonances is shifted (the Bloch-Siegert
shift) \cite{Goorden}. Thus, at fixing $\omega $ and $\Delta E$ and with
increasing the amplitude $x_{0}$ one should expect the (quasi-) periodical
behaviour due to the shift of the multiphoton resonances. Below we analyse
this issue in terms of the shift of the multiphoton resonances following
Ref. \cite{KrYak}. Alternatively the quasi-periodical behaviour of the
probability can be described in terms of the sequential Landau-Zener
transitions with taking into account the quantum-mechanical interference
between the transition events as in Ref. \cite{ShIF}.

Consider first, for simplicity, the case of the zero offset, $x_{off}=0$. In
this case the position of the resonances in the dependence of the occupation
probability on the system's parameters is defined by the following equation 
\cite{KrYak}:%
\begin{equation}
\frac{2\Delta }{\hbar \omega }\sqrt{1+q^{2}}E\left( \frac{q}{\sqrt{1+q^{2}}}%
\right) =\frac{\pi }{2}K,\text{ }K=1,3,5,...,  \label{from_KrYak}
\end{equation}%
where $E(k)=\dint\nolimits_{0}^{1}dx\frac{\sqrt{1-k^{2}x^{2}}}{\sqrt{1-x^{2}}%
}$ is the complete elliptic integral of the second kind and $q=x_{0}/\Delta $%
. The parameter $q$ is the key parameter of the problem; consider two
limiting cases: of the weak driving, $q\ll 1$, and of the very strong
driving, $q\gg 1$ (the term \textquotedblleft strong
driving\textquotedblright\ we reserve for the case $q\sim 1$); from Eq. (\ref%
{from_KrYak}) it follows:%
\begin{eqnarray}
\Delta E &=&K\cdot \hbar \omega \text{, }q\ll 1\text{,}  \label{multiphoton}
\\
\frac{4x_{0}}{\hbar \omega } &=&\pi K\text{, \ \ \ \ }q\gg 1.
\label{coherLZ}
\end{eqnarray}%
The first relation defines the multiphoton resonances, when the energy level
separation, $\Delta E=2\Delta \,$, is a multiple of a photon energy $\hbar
\omega $. The resonances, defined by Eq. (\ref{multiphoton}), can be
observed in the dependence of the occupation probability on $\omega $ or $%
\Delta $, but not in the dependence on $x_{0}$. In the second case the
resonances, defined by Eq. (\ref{coherLZ}), can be observed in the
dependence on $\omega $ or $x_{0}$, but not in the dependence on $\Delta $;
in this case equation (\ref{coherLZ}) also implies periodical (or
quasi-periodical) dependence on the parameter $\phi =4x_{0}/\hbar \omega $,
which was studied in Refs. \cite{ShIF} and \cite{ShKOK}. For the strong
driving, $q\sim 1$, the resonances are expected in dependencies on each of
the three parameters: $\omega $, $x_{0}$, and $\Delta $. Thus, we expect to
find in the regime of the strong driving features typical for the two
limiting cases: (i) quasi-periodic resonant dependence on $x_{0}$ and (ii)
the resonances to appear in the dependence on $\Delta $\ (with their
positions being dependent on $x_{0}$).

Consider the PBCPB \cite{Tinkham, italyanci, Zorin(02), FriedmanAverin,
Krech02, ShKOK} excited through the gate electrode. The PBCPB is the small
superconducting island, which is connected via two Josephson junctions
(characterized by energies $E_{J1,2}$ and phase differences $\delta _{1,2}$)
to the ring with low inductance $L$ (which is pierced by the magnetic flux $%
\Phi _{e}$) and via the capacitance $C_{g}$ to the gate with voltage $V_{g}$%
.\ The PBCPB is described by the Hamiltonian:%
\begin{equation}
\widehat{H}=\frac{\varepsilon _{J}}{2}\widehat{\sigma }%
_{x}-2E_{C}(1-n_{g}^{(0)}-n_{g}^{(1)}\sin \omega t)\widehat{\sigma }_{z},
\label{Ham_qb}
\end{equation}%
where the Coulomb energy of the island with the total capacitance $C_{tot}$\
is $E_{C}=e^{2}/2C_{tot}$; the effective Josephson energy is $\varepsilon
_{J}=\left( E_{J1}^{2}+E_{J2}^{2}+2E_{J1}E_{J2}\cos \delta \right) ^{1/2}$;
the total phase difference, $\delta =\delta _{1}+\delta _{2}$, approximately
equals to $2\pi \Phi _{e}/\Phi _{0}$; and the dimensionless gate voltage is $%
n_{g}(t)=n_{g}^{(0)}+n_{g}^{(1)}\sin \omega t=C_{g}V_{g}(t)/e$. The
Hamiltonian of the PBCPB (\ref{Ham_qb}) coincides with the introduced above
Hamiltonian (\ref{Ham_ShIF}) with the substitutions: $\Delta =\varepsilon
_{J}(\delta )/2$, $x_{off}=-2E_{C}(1-n_{g}^{(0)})$, and $%
x_{0}=2E_{C}n_{g}^{(1)}$.

Now the parameter $q$ is given by $q=\frac{4E_{C}}{\varepsilon _{J}}%
n_{g}^{(1)}$. Thus both limiting cases -- of weak and of very strong driving
-- described above, can in principle be realized in the PBCPB \cite{ShKOK},
where the domination of the Coulomb energy of a Cooper pair $4E_{C}$ over
the coupling energy $\varepsilon _{J}$ is assumed, $4E_{C}/\varepsilon
_{J}>1 $. In Ref. \cite{multiphoton} we have studied the case of weak
driving and here we study the case of strong driving, $q\sim 1$, in detail.

We will study the dependencies on $n_{g}^{(1)}$ and on $\delta $ to
demonstrate features (i) and (ii). The occupation probabilities of the PBCPB
are assumed to be probed by the tank circuit which is weakly coupled through
the mutual inductance $M$ to the PBCPB \cite{RandD,ili}. The average current 
$\left\langle \widehat{I}\right\rangle $ through the PBCPB is related to the
phase shift between the voltage and current $\alpha $, when the tank circuit
with the capacity $C_{T}$ and the inductunce $L_{T}$ is driven at the
resonant frequency $\omega _{T}=1/\sqrt{L_{T}C_{T}}$, as follows \cite%
{multiphoton}:%
\begin{equation}
\tan \alpha \simeq k^{2}QL\frac{2e}{\hbar }\frac{\partial \left\langle 
\widehat{I}\right\rangle }{\partial \delta },  \label{alpha}
\end{equation}%
where $Q^{-1}=\omega _{T}C_{T}R_{T}$, $k^{2}=M^{2}/(L\cdot L_{T})$. To
obtain the expectation value for the current in the qubit's ring, $%
\left\langle \widehat{I}\right\rangle =Sp\left( \widehat{\rho }\widehat{I}%
\right) $, we solve numerically the Bloch equations for the reduced density
matrix $\widehat{\rho }$, as we did in Ref. \cite{multiphoton}.

\begin{figure}[h]
\includegraphics[width=6cm]{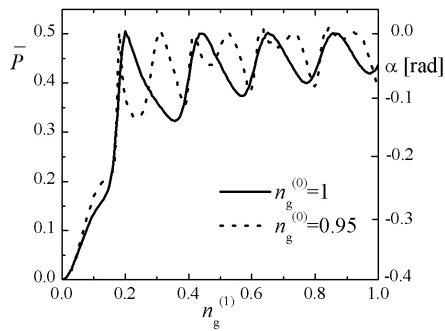}
\caption{Dependence of {}the time-averaged upper level occupation
probability $\overline{P}$ (left) and phase shift $\protect\alpha $ (right)
on the amplitude $n_{g}^{(1)}$ at $\protect\delta =\protect\pi $.}
\label{a_vs_ng1}
\end{figure}

In Fig. \ref{a_vs_ng1} we plot the time-averaged upper level occupation
probability $\overline{P}$ as a function of the amplitude $n_{g}^{(1)}$ at $%
\delta =\pi $ by making use of the solution of the Bloch equations. The case
of $n_{g}^{(0)}\neq 1$ (that is $x_{off}\neq 0$) differs from the case of $%
n_{g}^{(0)}=1$ ($x_{off}=0$) by the appearance of the additional peaks,
which was discussed in Refs.\ \cite{ShIF} and \cite{ShKOK}. We point out
that similar dependence, which illustrates the feature (i), in the case $%
x_{off}=0$ can be calculated alternatively by making use of other
approaches, namely with Eq. (13) from Ref. \cite{KrYak} and with Eq. (17)
from Ref. \cite{ShIF}. The numerical solution of the Bloch equations allows
us to overcome the restrictions of the analytical works: in Ref. \cite{KrYak}
neither decoherence nor $x_{off}\neq 0$ were taken into account while in
Ref. \cite{ShIF} the assumption of very strong driving was done, which, for
example, excludes the feature (ii). 

\begin{figure}[h]
\includegraphics[width=6cm]{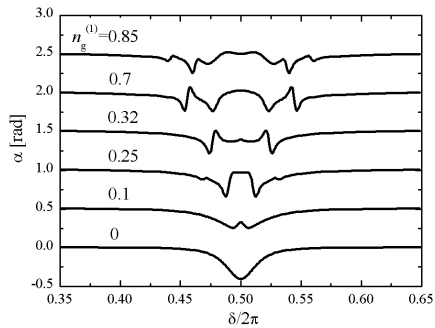}
\caption{The dependence of the phase shift $\protect\alpha $ on the total
phase difference $\protect\delta $ for different amplitudes $n_{g}^{(1)}$.
Upper curves are shifted vertically for clarity.}
\label{a_vs_d}
\end{figure}

Since at $\delta =\pi $ the phase shift $\alpha $ is proportional to the
time-averaged difference between the ground and excited state occupation
probabilities \cite{multiphoton}, $1-2\overline{P}$, Fig. \ref{a_vs_ng1}
presents also the dependence of $\alpha $ on $n_{g}^{(1)}$. In Fig. \ref%
{a_vs_d} the dependence of the phase shift $\alpha $ on the total phase
difference $\delta $ is plotted for different amplitudes $n_{g}^{(1)}$. Note
that, as it was explained in Ref. \cite{multiphoton}, the dependence of the
phase shift $\alpha $ on $\delta $ has hyperbolic-like character in the
vicinity of the resonances. The parameters of the system taken for the Figs. %
\ref{a_vs_ng1} and \ref{a_vs_d} are the following: $E_{J1}/E_{C}=4.5$, $%
E_{J2}/E_{C}=4$, $\hbar \omega /E_{C}=0.25$, $k^{2}Q\cdot 2e^{2}LE_{C}/\hbar
^{2}=0.01$; the temperature was considered to be zero (\textit{i.e.} much
less than $E_{J1}-E_{J2}$); the decoherence rates we considered to be the
functions of the energy level separation: $\Gamma _{1,2}\propto $ $\Delta
E(\delta )$ \cite{MShSh} (we have taken $\Gamma _{1,2}/E_{C}\sim 0.01$).
Such set of parameters was taken for the comparison with the experimental
results of Ref. \cite{VI}. 

In conclusion, we clarified from analytical consideration the qualitative
difference between the weak driving of a TLS and very strong driving. Then
the strongly driven PBCPB was studied. The numerical results (Figs. \ref%
{a_vs_ng1} and \ref{a_vs_d}) demonstrated that (i) the dependence of the
tank phase shift $\alpha $ on the amplitude $n_{g}^{(1)}$ at $\delta =\pi $
has resonant quasi-periodic character and (ii) the resonances appear in the
dependence on the phase difference $\delta $ as the amplitude-dependent
hyperbolic-like structures. We point out that the dependencies,
characterized by the features (i) and (ii), similar to Figs. \ref{a_vs_ng1}
and \ref{a_vs_d}, were observed experimentally \cite{VI}. And also similar
to Fig. \ref{a_vs_ng1} quasi-periodic dependence of the upper level
occupation probability on the driving (microwave) amplitude was observed in
the superconducting TLS based on a large Josephson-junction in Fig. 6 of
Ref. \cite{Martinis02}.

We thank the authors of Ref. \cite{VI} for communication of their
experimental results prior to the publication and E. Il'ichev, W. Krech and
V.I. Shnyrkov for fruitful discussions. The authors acknowledge the grant
\textquotedblleft Nanosystems, nanomaterials, and
nanotechnology\textquotedblright\ of the National Academy of Sciences of
Ukraine.

\textbf{Note added in proof.} During the preparation of the manuscript we
became aware that similar works on the strongly driven superconducting
systems were done \cite{finni, americanci}. The articles are given up to the
experimental and theoretical study of the interference fringes in the
strongly driven Cooper-pair transistor \cite{finni} and the flux qubit \cite%
{americanci}. The conclusions of the present work agree with the conclusions
of these articles.

\end{document}